
\def\service{T}
\catcode`\@=11
\def\unredoffs{\voffset=11mm \hoffset=0.5mm}

%
\newbox\leftpage \newdimen\fullhsize \newdimen\hstitle \newdimen\hsbody
\newdimen\hdim
\tolerance=400\pretolerance=800
%
%
\newif\ifsmall \smallfalse
\newif\ifdraft \draftfalse
\newif\iffrench \frenchfalse
\newif\ifeqnumerosimple \eqnumerosimplefalse
\nopagenumbers
\headline={\ifnum\pageno=1\hfill\else\hfil{\headrm\folio}\hfil\fi}
\def\draftstart{
\magnification=1200 \unredoffs\hsize=130mm\vsize=190mm
\hsbody=\hsize \hstitle=\hsize 
\nolabels
\iffrench
\dicof
\else
\dicoa
\fi
}

\font\elevrm=cmr9

\newdimen\chapskip
\font\twbf=cmssbx10 scaled 1200
\font\ssbx=cmssbx10

\font\caprm=cmr9
\font\capit=cmti9
\font\capbf=cmbx9
\font\capsl=cmsl9
\font\capmi=cmmi9
\font\capex=cmex9
\font\capsy=cmsy9
\chapskip=17.5mm
\def\makeheadline{\vbox to 0pt{\vskip-22.5pt
\line{\vbox to8.5pt{}\the\headline}\vss}\nointerlineskip}
\font\tbfi=cmmib10
\font\tenbi=cmmib7
\font\fivebi=cmmib5
\textfont4=\tbfi
\scriptfont4=\tenbi
\scriptscriptfont4=\fivebi
\font\headrm=cmr10

\font\eightrm=cmr6
\font\sixrm=cmr5
\font\eightmi=cmmi6
\font\sixmi=cmmi5
\font\eightsy=cmsy6
\font\sixsy=cmsy5
\font\eightbf=cmbx6
\font\sixbf=cmbx5
\skewchar\capmi='177 \skewchar\eightmi='177 \skewchar\sixmi='177
\skewchar\capsy='60 \skewchar\eightsy='60 \skewchar\sixsy='60

\def\elevenpoint{
\textfont0=\caprm \scriptfont0=\eightrm \scriptscriptfont0=\sixrm
\def\rm{\fam0\caprm}
\textfont1=\capmi \scriptfont1=\eightmi \scriptscriptfont1=\sixmi
\textfont2=\capsy \scriptfont2=\eightsy \scriptscriptfont2=\sixsy
\textfont3=\capex \scriptfont3=\capex \scriptscriptfont3=\capex
\textfont\itfam=\capit \def\it{\fam\itfam\capit} 
\textfont\slfam=\capsl  \def\sl{\fam\slfam\capsl} 
\textfont\bffam=\capbf \scriptfont\bffam=\eightbf
\scriptscriptfont\bffam=\sixbf
\def\bf{\fam\bffam\capbf} 
\textfont4=\tbfi \scriptfont4=\tenbi \scriptscriptfont4=\tenbi
\normalbaselineskip=13pt
\setbox\strutbox=\hbox{\vrule height9.5pt depth3.9pt width0pt}
\let\big=\elevenbig \normalbaselines \rm}

\catcode`\@=11

\font\tenmsa=msam10
\font\sevenmsa=msam7
\font\fivemsa=msam5
\font\tenmsb=msbm10
\font\sevenmsb=msbm7
\font\fivemsb=msbm5
\newfam\msafam
\newfam\msbfam
\textfont\msafam=\tenmsa  \scriptfont\msafam=\sevenmsa
  \scriptscriptfont\msafam=\fivemsa
\textfont\msbfam=\tenmsb  \scriptfont\msbfam=\sevenmsb
  \scriptscriptfont\msbfam=\fivemsb

\def\hexnumber@#1{\ifcase#1 0\or1\or2\or3\or4\or5\or6\or7\or8\or9\or
	A\or B\or C\or D\or E\or F\fi }

\font\teneuf=eufm10
\font\seveneuf=eufm7
\font\fiveeuf=eufm5
\newfam\euffam
\textfont\euffam=\teneuf
\scriptfont\euffam=\seveneuf
\scriptscriptfont\euffam=\fiveeuf
\def\frak{\ifmmode\let\next\frak@\else
 \def\next{\Err@{Use \string\frak\space only in math mode}}\fi\next}
\def\goth{\ifmmode\let\next\frak@\else
 \def\next{\Err@{Use \string\goth\space only in math mode}}\fi\next}
\def\frak@#1{{\frak@@{#1}}}
\def\frak@@#1{\fam\euffam#1}

\edef\msa@{\hexnumber@\msafam}
\edef\msb@{\hexnumber@\msbfam}

\def\Bbb{\ifmmode\let\next\Bbb@\else
 \def\next{\errmessage{Use \string\Bbb\space only in math mode}}\fi\next}
\def\Bbb@#1{{\Bbb@@{#1}}}
\def\Bbb@@#1{\fam\msbfam#1}

\catcode`\@=12
\def\sla#1{\mkern-1.5mu\raise0.4pt\hbox{$\not$}\mkern1.2mu #1\mkern 0.7mu}
\def\Dbar{\mkern-1.5mu\raise0.4pt\hbox{$\not$}\mkern-.1mu {\rm D}\mkern.1mu}
\def\Abar{\mkern1.mu\raise0.4pt\hbox{$\not$}\mkern-1.3mu A\mkern.1mu}
\def\dicof{
\gdef\Resume{RESUME}
\gdef\Toc{Table des mati\`eres}
\gdef\soumisa{Soumis \`a:}
}
\def\dicoa{
\gdef\Resume{ABSTRACT}
\gdef\Toc{Table of Contents}
\gdef\soumisa{Submitted to}
}

\def\uniset{\rlap{\elevrm 1}\kern.15em 1}
\def\bkR{{\rm I\kern-.17em R}}
\def\bkC{{\rm \kern.24em
            \vrule width.05em height1.4ex depth-.05ex
            \kern-.26em C}}

\def\frac#1#2{{\textstyle{#1\over#2}}}

\def\leaderfill{\leaders\hbox to 1em{\hss.\hss}\hfill}
\def\saclay{\if S\service \spec \else \spht \fi}
\def\spht{
\centerline{CEA, Service de Physique Th\'eorique, CE-Saclay}
\centerline{F-91191 Gif-sur-Yvette Cedex, FRANCE}}
\def\spec{
\centerline{CEA/DSM/DRECAM/Service de Physique de l'Etat Condens\'e}
\centerline{CE Saclay, F-91191 Gif-sur-Yvette Cedex, FRANCE}}
\def\logo{
\if S\service 
\font\sstw=cmss10 scaled 1200
\font\ssx=cmss8
\vtop{\hsize 9cm
{\sstw {\twbf P}hysique de l'{\twbf E}tat {\twbf C}ondens\'e \par}
\ssx SPEC -- DRECAM -- DSM\par
\vskip 0.5mm
\sstw CEA -- Saclay \par
}
\else 
\vtop{\hsize 9cm
}
\fi }
\catcode`\@=11
\def\deqalignno#1{\displ@y\tabskip\centering \halign to
\displaywidth{\hfil$\displaystyle{##}$\tabskip0pt&$\displaystyle{{}##}$
\hfil\tabskip0pt &\quad
\hfil$\displaystyle{##}$\tabskip0pt&$\displaystyle{{}##}$
\hfil\tabskip\centering& \llap{$##$}\tabskip0pt \crcr #1 \crcr}}
\def\deqalign#1{\null\,\vcenter{\openup\jot\m@th\ialign{
\strut\hfil$\displaystyle{##}$&$\displaystyle{{}##}$\hfil
&&\quad\strut\hfil$\displaystyle{##}$&$\displaystyle{{}##}$
\hfil\crcr#1\crcr}}\,}
\openin 1=\jobname.sym
\ifeof 1\closein1\message{<< (\jobname.sym DOES NOT EXIST) >>}\else%
\input\jobname.sym\closein 1\fi
\newcount\nosection
\newcount\nosubsection
\newcount\neqno
\newcount\notenumber
\newcount\figno
\newcount\tabno
\def\content{\jobname.toc}
\def\symbols{\jobname.sym}
\newwrite\toc
\newwrite\sym
\def\authorname#1{\centerline{\bf #1}\smallskip}
\def\address#1{ #1\medskip}
\newdimen\hulp
\def\maketitle#1{
\edef\oneliner##1{\centerline{##1}}
\edef\twoliner##1{\vbox{\parindent=0pt\leftskip=0pt plus 1fill\rightskip=0pt
plus 1fill
                     \parfillskip=0pt\relax##1}}
\setbox0=\vbox{#1}\hulp=0.5\hsize
                 \ifdim\wd0<\hulp\oneliner{#1}\else
                 \twoliner{#1}\fi}
\def\pacs#1{{\bf PACS numbers:} #1\par}
\def\submitted#1{{\it {\soumisa} #1}\par}
\def\title#1{\gdef\titlename{#1}
\maketitle{
\twbf
{\titlename}}
\vskip3truemm\vfill
\nosection=0
\neqno=0
\notenumber=0
\figno=1
\tabno=1
\def\prefix{}
\def\eqprefix{}
\mark{\the\nosection}
\message{#1}
\immediate\openout\sym=\symbols
}
\def\preprint#1{\vglue-10mm
\line{ \logo \hfill {#1} }\vglue 20mm\vfill}
\def\abstract{\vfill\centerline{\Resume} \smallskip \begingroup\narrower
\elevenpoint\baselineskip10pt}
\def\endabstract{\par\endgroup \bigskip}
\def\mktoc{\centerline{\bf \Toc} \medskip\caprm
\parindent=2em
\openin 1=\jobname.toc
\ifeof 1\closein1\message{<< (\jobname.toc DOES NOT EXIST. TeX again)>>}%
\else\input\jobname.toc\closein 1\fi
 \bigskip}
\def\section#1\par{\vskip0pt plus.1\vsize\penalty-100\vskip0pt plus-.1
\vsize\bigskip\vskip\parskip
\message{ #1}
\ifnum\nosection=0\immediate\openout\toc=\content%
\edef\ecrire{\write\toc{\par\noindent{\ssbx\ \titlename}
\string\leaderfill{\noexpand\number\pageno}}}\ecrire\fi
\advance\nosection by 1\nosubsection=0
\ifeqnumerosimple
\else \xdef\eqprefix{\prefix\the\nosection.}\neqno=0\fi
\vbox{\noindent\bf\prefix\the\nosection\ #1}
\mark{\the\nosection}\bigskip\noindent
\xdef\ecrire{\write\toc{\string\par\string\item{\prefix\the\nosection}
#1
\string\leaderfill {\noexpand\number\pageno}}}\ecrire}

\def\appendix#1#2\par{\bigbreak\nosection=0
\notenumber=0
\neqno=0
\def\prefix{A}
\mark{\the\nosection}
\message{\appendixname}
\leftline{\ssbx APPENDIX}
\leftline{\ssbx\uppercase\expandafter{#1}}
\leftline{\ssbx\uppercase\expandafter{#2}}
\bigskip\noindent\nonfrenchspacing
\edef\ecrire{\write\toc{\par\noindent{{\ssbx A}\
{\ssbx#1\ #2}}\string\leaderfill{\noexpand\number\pageno}}}\ecrire}%

\def\subsection#1\par {\vskip0pt plus.05\vsize\penalty-100\vskip0pt
plus-.05\vsize\bigskip\vskip\parskip\advance\nosubsection by 1
\vbox{\noindent\it\prefix\the\nosection.\the\nosubsection\
\it #1}\smallskip\noindent
\edef\ecrire{\write\toc{\string\par\string\itemitem
{\prefix\the\nosection.\the\nosubsection} {#1}
\string\leaderfill{\noexpand\number\pageno}}}\ecrire
}
\def\note #1{\advance\notenumber by 1
\footnote{$^{\the\notenumber}$}{\sevenrm #1}}

\def\nolabels{\def\wrlabel##1{}\def\eqlabel##1{}\def\reflabel##1{}}
\def\writelabels{\def\wrlabel##1{\leavevmode\vadjust{\rlap{\smash%
{\line{{\escapechar=` \hfill\rlap{\sevenrm\hskip.03in\string##1}}}}}}}%
\def\eqlabel##1{{\escapechar-1\rlap{\sevenrm\hskip.05in\string##1}}}%
\def\reflabel##1{\noexpand\llap{\noexpand\sevenrm\string\string\string##1}}}
\global\newcount\refno \global\refno=1
\newwrite\rfile
\def\ref{[\the\refno]\nref}
\def\nref#1{\xdef#1{[\the\refno]}\writedef{#1\leftbracket#1}%
\ifnum\refno=1\immediate\openout\rfile=\jobname.ref\fi
\global\advance\refno by1\chardef\wfile=\rfile\immediate
\write\rfile{\noexpand\item{#1\ }\reflabel{#1\hskip.31in}\pctsign}\findarg}
\def\findarg#1#{\begingroup\obeylines\newlinechar=`\^^M\pass@rg}
{\obeylines\gdef\pass@rg#1{\writ@line\relax #1^^M\hbox{}^^M}%
\gdef\writ@line#1^^M{\expandafter\toks0\expandafter{\striprel@x #1}%
\edef\next{\the\toks0}\ifx\next\em@rk\let\next=\endgroup\else\ifx\next\empty%
\else\immediate\write\wfile{\the\toks0}\fi\let\next=\writ@line\fi\next\relax}}
\def\striprel@x#1{}
\def\em@rk{\hbox{}}

\def\addref#1{\immediate\write\rfile{\noexpand\item{}#1}} 
\def\listrefs{
\ifnum\refno=1 \else
\immediate\closeout\rfile\writestoppt\baselineskip=14pt%
\vskip0pt plus.1\vsize\penalty-100\vskip0pt plus-.1
\vsize\bigskip\vskip\parskip\centerline{{\bf References}}\bigskip%
{\frenchspacing%
\parindent=20pt\escapechar=` \input \jobname.ref\vfill\eject}%
\nonfrenchspacing
\fi}
\def\startrefs#1{\immediate\openout\rfile=\jobname.ref\refno=#1}
\def\xref{\expandafter\xr@f}\def\xr@f[#1]{#1}
\def\refs#1{[\r@fs #1{\hbox{}}]}
\def\r@fs#1{\ifx\und@fined#1\message{reflabel \string#1 is undefined.}%
\xdef#1{(?.?)}\fi \edef\next{#1}\ifx\next\em@rk\def\next{}%
\else\ifx\next#1\xref#1\else#1\fi\let\next=\r@fs\fi\next}
%
\newwrite\lfile
{\escapechar-1\xdef\pctsign{\string\%}\xdef\leftbracket{\string\{}
\xdef\rightbracket{\string\}}}

\def\writestop{\def\writestoppt{\immediate\write\lfile{\string\pageno%
\the\pageno\string\startrefs\leftbracket\the\refno\rightbracket%
\string\def\string\secsym\leftbracket\secsym\rightbracket%
\string\secno\the\secno\string\meqno\the\meqno}\immediate\closeout\lfile}}
\def\writestoppt{}\def\writedef#1{}
\def\eqnn{\global\advance\neqno by 1 \ifinner\relax\else%
\eqno\fi(\eqprefix\the\neqno)}
%
\def\eqnd#1{\global\advance\neqno by 1 \ifinner\relax\else%
\eqno\fi(\eqprefix\the\neqno)\eqlabel#1
{\xdef#1{($\eqprefix\the\neqno$)}}
\edef\ewrite{\write\sym{\string\def\string#1{($\eqprefix%
\the\neqno$)}}%
}\ewrite%
}
%
\def\eqna#1{\wrlabel#1\global\advance\neqno by1
{\xdef #1##1{\hbox{$(\eqprefix\the\neqno##1)$}}}
\edef\ewrite{\write\sym{\string\def\string#1{($\eqprefix%
\the\neqno$)}}%
}\ewrite%
}
\def\em@rk{\hbox{}}
\def\xeqn{\expandafter\xe@n}\def\xe@n(#1){#1}
\def\xeqna#1{\expandafter\xe@na#1}\def\xe@na\hbox#1{\xe@nap #1}
\def\xe@nap$(#1)${\hbox{$#1$}}
\def\eqns#1{(\e@ns #1{\hbox{}})}
\def\e@ns#1{\ifx\und@fined#1\message{eqnlabel \string#1 is undefined.}%
\xdef#1{(?.?)}\fi \edef\next{#1}\ifx\next\em@rk\def\next{}%
\else\ifx\next#1\xeqn#1\else\def\n@xt{#1}\ifx\n@xt\next#1\else\xeqna#1\fi
\fi\let\next=\e@ns\fi\next}
\def\fig{fig.~\the\figno\nfig}
\def\nfig#1{\xdef#1{\the\figno}%
\immediate\write\sym{\string\def\string#1{\the\figno}}%
\global\advance\figno by1}%
\def\xfig{\expandafter\xf@g}\def\xf@g fig.\penalty\@M\ {}%
\def\figs#1{figs.~\f@gs #1{\hbox{}}}%
\def\f@gs#1{\edef\next{#1}\ifx\next\em@rk\def\next{}\else%
\ifx\next#1\xfig #1\else#1\fi\let\next=\f@gs\fi\next}%
\long\def\figure#1#2#3{\midinsert
#2\par
{\elevenpoint
\setbox1=\hbox{#3}
\ifdim\wd1=0pt\centerline{{\bf Figure\ #1}\hskip7.5mm}%
\else\setbox0=\hbox{{\bf Figure #1}\quad#3\hskip7mm}
\ifdim\wd0>\hsize{\narrower\noindent\unhbox0\par}\else\centerline{\box0}\fi
\fi}
\wrlabel#1\par
\endinsert}
\def\tab{table~\uppercase\expandafter{\romannumeral\the\tabno}\ntab}
\def\ntab#1{\xdef#1{\the\tabno}
\immediate\write\sym{\string\def\string#1{\the\tabno}}
\global\advance\tabno by1}
\long\def\table#1#2#3{\topinsert
#2\par
{\elevenpoint
\setbox1=\hbox{#3}
\ifdim\wd1=0pt\centerline{{\bf Table
\uppercase\expandafter{\romannumeral#1}}\hskip7.5mm}%
\else\setbox0=\hbox{{\bf Table
\uppercase\expandafter{\romannumeral#1}}\quad#3\hskip7mm}
\ifdim\wd0>\hsize{\narrower\noindent\unhbox0\par}\else\centerline{\box0}\fi
\fi}
\wrlabel#1\par
\endinsert}
\catcode`@=12
\def\draftend{\immediate\closeout\sym\immediate\closeout\toc
}
\draftstart
\preprint{T94/134}
\title{Random field Ising model: dimensional reduction or spin-glass phase?}
\authorname{C. De Dominicis,
H. Orland}
\address{\saclay}
\authorname{T. Temesvari}
\address{\centerline{Institute for Theoretical Physics}
\centerline{E\"otv\"os University, H-1088 Budapest, HUNGARY}
}
\abstract
The stability of the random field Ising model (RFIM) against spin
glass (SG) fluctuations, as investigated by M\'ezard and Young, is naturally
expressed via Legendre transforms, stability being then associated with the
non-negativeness of eigenvalues of the inverse of a generalized SG
susceptibility matrix. It is found that the signal for the
occurrence of the SG transition will manifest itself in free-energy
{\sl fluctuations\/} only, and not in the free energy itself.

Eigenvalues of the inverse SG susceptibility matrix is then approached by the
Rayleigh Ritz method which provides an upper bound. Coming from the
paramagnetic phase {\sl on the Curie line,\/} one is able to use a
virial-like relationship generated by scaling the {\sl single\/} unit length
$ (D<6; $ in higher dimension a new length sets in, the inverse momentum cut
off). Instability towards a SG phase being probed on pairs of {\sl distinct\/}
replicas, it follows that, despite the repulsive coupling of the RFIM the
effective pair coupling is {\sl attractive\/} (at least for small values of
the parameter $ g\bar \Delta , $ $ g $ the coupling and $ \bar \Delta $ the
effective random field fluctuation). As a result, \lq\lq bound states\rq\rq\
associated with replica pairs (negative eigenvalues) provide the instability
signature. {\sl Away from the Curie line\/}, the attraction is damped out till
the SG transition line is reached and paramagnetism restored. In $ D<6, $ the
SG transition always precedes the ferromagnetic one, thus the domain in
dimension where standard dimensional reduction would apply (on the Curie
line) shrinks to zero.
\endabstract
\vfill
\submitted{J. de Physique I}
\eject
\eject
\def\m@th{\mathsurround=0pt}
\newif\ifdtpt
\def\displ@y{\openup1\jot\m@th
    \everycr{\noalign{\ifdtpt\dt@pfalse
    \vskip-\lineskiplimit \vskip\normallineskiplimit
    \else \penalty\interdisplaylinepenalty \fi}}}
\def\eqalignc#1{\,\vcenter{\openup1\jot\m@th
                \ialign{\strut\hfil$\displaystyle{##}$\hfil&
                              \hfil$\displaystyle{{}##}$\hfil&
                              \hfil$\displaystyle{{}##}$\hfil&
                              \hfil$\displaystyle{{}##}$\hfil&
                              \hfil$\displaystyle{{}##}$\hfil\crcr#1\crcr}}\,}
\def\eqalignnoc#1{\displ@y\tabskip\centering \halign to \displaywidth{
                  \hfil$\displaystyle{##}$\hfil\tabskip=0pt &
                  \hfil$\displaystyle{{}##}$\hfil\tabskip=0pt &
                  \hfil$\displaystyle{{}##}$\hfil\tabskip=0pt &
                  \hfil$\displaystyle{{}##}$\hfil\tabskip=0pt &
                  \hfil$\displaystyle{{}##}$\hfil\tabskip\centering &
                  \llap{$##$}\tabskip=0pt \crcr#1\crcr}}
\def\leqalignnoc#1{\displ@y\tabskip\centering \halign to \displaywidth{
                  \hfil$\displaystyle{##}$\hfil\tabskip=0pt &
                  \hfil$\displaystyle{{}##}$\hfil\tabskip=0pt &
                  \hfil$\displaystyle{{}##}$\hfil\tabskip=0pt &
                  \hfil$\displaystyle{{}##}$\hfil\tabskip=0pt &
                  \hfil$\displaystyle{{}##}$\hfil\tabskip\centering &
                  \kern-\displaywidth\rlap{$##$}\tabskip=\displaywidth
                  \crcr#1\crcr}}
\def\dasharrowfill{$\mathsurround=0pt \mathord- \mkern-6mu
    \cleaders\hbox{$\mkern-2mu\mathord-\mkern-2mu$}\hfill
    \mkern-6mu \mathord-$}

\def\doublelow#1{\,\vtop{\ialign{\hfil$##$\hfil\crcr
                 \mathstrut #1 \crcr}}\,}
\def\charlvmidlw#1#2{\,\vtop{\ialign{##\crcr
      #1\crcr\noalign{\kern1pt\nointerlineskip}
      $\hfil#2\hfil$\crcr}}\,}
\def\charlvlowlw#1#2{\,\vtop{\ialign{##\crcr
      $\hfil#1\hfil$\crcr\noalign{\kern1pt\nointerlineskip}
      #2\crcr}}\,}
\def\charlvmidup#1#2{\,\vbox{\ialign{##\crcr
      $\hfil#1\hfil$\crcr\noalign{\kern1pt\nointerlineskip}
      #2\crcr}}\,}
\def\charlvupup#1#2{\,\vbox{\ialign{##\crcr
      #1\crcr\noalign{\kern1pt\nointerlineskip}
      $\hfil#2\hfil$\crcr}}\,}
 
\def\vspce{\kern4pt} \def\hspce{\kern4pt}    

\def\emptybox{\vbox{\kern.7ex\hbox{\kern.5em}\kern.7ex}}
 \font\sevmi  = cmmi7              
    \skewchar\sevmi ='177
 \font\fivmi  = cmmi5              
    \skewchar\fivmi ='177
\font\tenmib=cmmib10
\newfam\bfmitfam

\textfont\bfmitfam=\tenmib
\scriptfont\bfmitfam=\sevmi
\scriptscriptfont\bfmitfam=\fivmi

\def\mathcedilla{\vtop{\hbox{c}{\kern0pt\nointerlineskip}
	         {\hbox{$\mkern-2mu \mathchar"0018\mkern-2mu$}}}}

\mathchardef\gq="0060
\mathchardef\dq="0027
\mathchardef\ssmath="19
\mathchardef\aemath="1A
\mathchardef\oemath="1B
\mathchardef\omath="1C
\mathchardef\AEmath="1D
\mathchardef\OEmath="1E
\mathchardef\Omath="1F
\mathchardef\imath="10
\mathchardef\fmath="0166
\mathchardef\gmath="0167
\mathchardef\vmath="0176

\def\twodot{.\kern-0.1em.}

\def\paral{\mathrel{/\kern-.25em/}}
\def\grlo{\mathrel{\hbox{\lower.2ex\hbox{\rlap{$>$}\raise1ex\hbox{$<$}}}}}
\def\logr{\mathrel{\hbox{\lower.2ex\hbox{\rlap{$<$}\raise1ex\hbox{$>$}}}}}
\def\greq{\mathrel{\hbox{\lower1ex\hbox{\rlap{$=$}\raise1.2ex\hbox{$>$}}}}}
\def\loeq{\mathrel{\hbox{\lower1ex\hbox{\rlap{$=$}\raise1.2ex\hbox{$<$}}}}}
\def\grsim{\mathrel{\hbox{\lower1ex\hbox{\rlap{$\sim$}\raise1ex\hbox{$>$}}}}}
\def\losim{\mathrel{\hbox{\lower1ex\hbox{\rlap{$\sim$}\raise1ex\hbox{$<$}}}}}
\font\ninerm=cmr9
\def\uniset{\rlap{\ninerm 1}\kern.15em 1}

\def\emptysq{\mathbin{\vbox{\hrule\hbox{\vrule height1ex \kern.5em
                            \vrule height1ex}\hrule}}}
\def\emptyrect{\mathbin{\vbox{\hrule\hbox{\vrule height1ex \kern1em
                              \vrule height1ex}\hrule}}}
\def\rightonleftarrow{\mathrel{\hbox{\raise.5ex\hbox{$\rightarrow$}\ignorespaces
                                   \lower.5ex\hbox{\llap{$\leftarrow$}}}}}
\def\leftonrightarrow{\mathrel{\hbox{\raise.5ex\hbox{$\leftarrow$}\ignorespaces
                                   \lower.5ex\hbox{\llap{$\rightarrow$}}}}}

\def\bkB{{\rm I\kern-.17em B}}
\def\bkC{{\rm \kern.24em
            \vrule width.05em height1.4ex depth-.05ex
            \kern-.26em C}}
\def\bkD{{\rm I\kern-.17em D}}
\def\bkE{{\rm I\kern-.17em E}}
\def\bkF{{\rm I\kern-.17em F}}
\def\bkG{{\rm \kern.24em
            \vrule width.05em height1.4ex depth-.05ex
            \kern-.26em G}}
\def\bkH{{\rm I\kern-.22em H}}
\def\bkI{{\rm I\kern-.22em I}}
\def\bkJ{{\rm \kern.19em
            \vrule width.02em height1.5ex depth0ex
            \kern-.20em J}}
\def\bkK{{\rm I\kern-.22em K}}
\def\bkL{{\rm I\kern-.17em L}}
\def\bkM{{\rm I\kern-.22em M}}
\def\bkN{{\rm I\kern-.20em N}}
\def\bkO{{\rm \kern.24em
            \vrule width.05em height1.4ex depth-.05ex
            \kern-.26em O}}
\def\bkP{{\rm I\kern-.17em P}}
\def\bkQ{{\rm \kern.24em
            \vrule width.05em height1.4ex depth-.05ex
            \kern-.26em Q}}
\def\bkR{{\rm I\kern-.17em R}}
\def\bkT{{\rm \kern.24em
            \vrule width.02em height1.5ex depth 0ex
            \kern-.27em T}}
\def\bkU{{\rm \kern.30em
            \vrule width.02em height1.47ex depth-.05ex
            \kern-.32em U}}
\def\bkZ{{\rm Z\kern-.32em Z}}

\vskip 20pt
\centerline{{\bf RANDOM FIELD ISING MODEL: DIMENSIONAL REDUCTION}}
\centerline{{\bf OR SPIN-GLASS PHASE?}}
\vskip 17pt
\centerline{by}
\vskip 15pt
\centerline{{\bf C. De Dominicis, H. Orland}}
\vskip 12pt
\centerline{CEA, Service de Physique Th\'eorique, CE-Saclay}
\centerline{F-91191 Gif-sur-Yvette Cedex, FRANCE}
\vskip 15pt
\centerline{and}
\vskip 15pt
\centerline{{\bf T. Temesvari}}
\vskip 12pt
\centerline{Institute for Theoretical Physics, E\"otv\"os University}
\centerline{H-1088 Budapest, HUNGARY}
\vskip 24pt
{\bf ABSTRACT}

The stability of the random field Ising model (RFIM) against spin
glass (SG) fluctuations, as investigated by M\'ezard and Young, is naturally
expressed via Legendre transforms, stability being then associated with the
non-negativeness of eigenvalues of the inverse of a generalized SG
susceptibility matrix. It is interesting to note that the signal for the
occurrence of the SG transition will manifest itself in free-energy
{\sl fluctuations\/} only, i.e., in the $ n $-replica formulation, in terms of
order $ n^2 $
and higher and not in the free energy itself (when the SG phase sets in, all
terms of order $ n^2 $ or higher, become linear and start contributing to
the free energy, a behavior contrasting with the standard SG).

Eigenvalues of the inverse SG susceptibility matrix can be approached by the
Rayleigh Ritz method which provides an upper bound. Coming from the
paramagnetic phase {\sl on the Curie line,\/} one is then able to use a
virial-like
relationship generated by scaling the {\sl single\/} unit length $ (D<6; $ in
higher
dimension a new length sets in, the inverse momentum cut off). Instability
towards a SG phase being probed on pairs of {\sl distinct\/} replicas, it
follows
that, despite the repulsive coupling of the RFIM the effective pair coupling
is {\sl attractive\/} (at least for small values of the parameter $ g\bar
\Delta , $ $ g $ the
coupling and $ \bar \Delta $ the effective random field fluctuation).
As a result, \lq\lq bound states\rq\rq\
associated with replica pairs (negative eigenvalues) provide the instability
signature. {\sl Away from the Curie line\/}, the attraction is damped out till
the SG
transition line is reached and paramagnetism restored. In $ D<6, $ the SG
transition always precedes the ferromagnetic one, thus the domain in
dimension where standard dimensional reduction would apply, shrinks to zero.
\vfill\eject
\centerline{{\bf MODELE D'ISING EN CHAMP ALEATOIRE: REDUCTION DIMENSIONNELLE}}
\centerline{{\bf OU PHASE VERRE DE SPIN?}}
\vskip 17pt
\centerline{par}
\vskip 15pt
\centerline{C. De Dominicis, H. Orland}
\vskip 12pt
\centerline{CEA, Service de Physique Th\'eorique, CE-Saclay}
\centerline{F-91191 Gif-sur-Yvette Cedex, FRANCE}
\vskip 15pt
\centerline{and}
\vskip 15pt
\centerline{T. Temesvari}
\vskip 12pt
\centerline{Institute for Theoretical Physics, E\"otv\"os University}
\centerline{H-1088 Budapest, HUNGARY}
\vskip 24pt
{\bf RESUME}

On examine la stabilit\'e du mod\`ele d'Ising en champ al\'eatoire (MICA) par
rapport aux fluctuations verre de spin (VS) comme chez M\'ezard et Young.
Cette
stabilit\'e s'exprime naturellement via des transformations de Legendre o\`u
elle
est associ\'ee \`a la non-n\'egativit\'e des valeurs propres de la matrice
inverse
d'une susceptibilit\'e VS g\'en\'eralis\'ee. Il est int\'eressant de noter que
le
signal de la transition VS n'appara\^\i t que dans les {\sl fluctuations\/} de
l'\'energie
libre, c'est-\`a-dire, dans la formulation en $ n $-r\'epliques, que dans les
termes
d'ordre $ n^2 $ ou plus \'elev\'e et non dans l'\'energie libre elle-m\^eme
(\`a la
transition VS, les termes d'ordre $ n^2 $ ou plus \'elev\'e, se transforment
tous en
termes lin\'eaire en $ n $ et contribuent alors \`a l'\'energie libre, un
comportement
qui contraste d'avec celui du VS standard).

Les valeurs propres de la matrice inverse de la susceptibilit\'e VS peuvent
s'estimer par la m\'ethode de Rayleigh-Ritz qui fournit une borne
sup\'erieure.
{\sl Sur la ligne de Curie,\/} en s'approchant \`a partir de la phase
paramagn\'etique,
on peut utiliser alors une relation du type \lq\lq viriel\rq\rq\ obtenue en
dilatant
l'{\sl unique\/} \'echelle des longueurs $ (D<6; $ en dimension sup\'erieure
une deuxi\`eme
\'echelle s'introduit, l'inverse du moment de coupure). L'instabilit\'e VS se
manifestant sur les corr\'elations de paires de r\'epliques {\sl distinctes\/}
il
s'ensuit que, malgr\'e le couplage r\'epulsif du MICA, le couplage effectif
d'une
paire est {\sl attractif\/} (du moins pour les faibles valeurs du param\`etre
$ g\bar \Delta , $ o\`u
$ g $ est le couplage et $ \bar \Delta $ la fluctuation effective du champ
al\'eatoire). Il en r\'esulte
des \lq\lq\'etats li\'es\rq\rq\ associ\'es aux paires de r\'epliques (valeurs
propres n\'egatives)
qui sont la signature de l'instabilit\'e. Quand on {\sl s'\'eloigne de la
ligne de
Curie,\/} l'attraction de paires est amortie, jusqu'\`a la ligne de transition
VS
o\`u le paramagn\'etisme appara\^\i t. En dimension $ D<6, $ la transition VS
pr\'ec\`ede
toujours la ferromagn\'etique, si bien que l'intervalle de dimension o\`u la
r\'eduction dimensionnelle pourrait s'appliquer, se r\'eduit \`a z\'ero.
\vfill\eject
After nearly twenty years of intense activity, there is yet no
consensus on the critical behavior of random field systems (for recent
reviews see $ [1,2]). $

A blatant contradiction arose when a calculation to all orders in
perturbations$ ^{[3,4]}, $ later supported by a non perturbative approach$
^{[5]}, $ established dimensional reduction (between the RFIM in
dimension $ D $ and the pure Ising system in $ D-\theta ) $ both for
hyperscaling
relationships between critical exponents and for the exponents themselves as
a function of $ D. $ With $ \theta =2 $ this was predicting a lower critical
dimension $ D_{\ell} =3 $
for the existence of a ferromagnetic phase, in contradiction with an early
Imry-Ma$ ^{[6]} $ argument predicting $ D_{\ell} =2 $ (later supported by
rigorous work of
Imbrie$ ^{[7]}, $ proving the existence of ferromagnetism in $ D=3). $

Despite the fact that $ D_{\ell} =2 $ is now widely accepted, there remains
the question
of down to which dimension are the resummed perturbation results valid,
and what happens below that dimension.

Meanwhile several groups have proposed the existence of a glassy phase sector
in the $ \Delta ,T $ plane $ (\Delta $ is the width of the random field
gaussian distribution
and $ T $ the temperature)
out of numerical studies$ ^{[8-12]} $ or from analytical work, extending to
random
field systems$ ^{[13-15]} $ the techniques of replica symmetry breaking$
^{[16]} $ (RSB).
In particular M\'ezard and Young$ ^{[14]} $ have used as a control parameter
the
number of components $ m $ (as in Bray$ ^{[17]} $ self consistent screening
approximation) and written out explicit self consistent equations for the
exponents.

Here, we follow the most straightforward approach to analyze properties of
the RFIM, i.e. we mimick what is done in the pure system to describe the
paramagnetic and condensed (ferromagnetic) phases.

We know since the work of Yvon$ ^{[18]} $ that the appropriate way to have
access to
the condensed phase is to replace the expansion in the local field $ H_i, $ by
one
in the local magnetization $ M_i, $ through a Legendre transform. The Jacobian
of
the transform $ {\rm det} \left(\partial M_i/\partial H_j \right) $ vanishes
at the transition (with the lowest
eigenvalue of the matrix $ \partial M_i/\partial H_j), $ displaying the
non-equivalence of the $ H_i $
and the $ M_i $ expansions.

Likewise here we consider the RFIM described by an effective hamiltonian with
an external field $ \Delta $ and perform the appropriate Legendre transform to
the
conjugate observable. Again the lowest eigenvalue of the jacobian matrix
yields the locus of the singularities of the associated susceptibility, here
the SG susceptibility, i.e. the line of the SG transition.

In Section 1-2, we recall the perturbation expansion and effective
hamiltonian for the RFIM. In Section 3, the Legendre transform is effected
yielding stationarity conditions and eigenvalue equations for the SG
transition. In Section 4 we study the transition and show that it manifests
itself in the free-energy fluctuation (as contrasted with the standard SG).
Section 5 is devoted to a study of the phase diagram using the Rayleigh-Ritz
variational method$ ^{[19]} $ and we conclude in Section 6, where our results
are
summarized.
\vskip 24pt
\vfill\eject
\noindent {\bf 1. THE PURE ISING SYSTEM}

Consider the pure Ising hamiltonian, in its soft spin version,
$$ {\cal H}={1 \over 2} \sum^{ }_ p \left(t_0+p^2 \right)\varphi (p)\varphi
(-p)+{g \over 4!} \sum^{ }_ j\varphi^ 4_j- \sum^{ }_ jH_j\varphi_ j \eqno
(1.1) $$
and
$$ W \left\{ H_j \right\} = {\rm ln} \int^{ }_{ } D\varphi \ {\rm exp} \
-{\cal H}\{ \varphi\} \equiv {\rm ln} \ Z \left\{ H_j \right\} \eqno (1.2) $$
One may describe the system by expanding in $ H; $ or via a Legendre transform
$$ W \left\{ H_j \right\} =-\Gamma \left\{ M_j \right\} + \sum^{ }_ jH_jM_j
\eqno (1.3) $$
where the magnetization $ M_i $ is
$$ M_j={\partial W \over \partial H_j}= \left\langle \varphi_ j \right\rangle
\eqno (1.4) $$
and the Legendre transform $ \Gamma $ satisfies
$$ H_j={\partial \Gamma \over \partial M_j}\ , \eqno (1.5) $$
by expanding in $ M. $ The bracket in (1.4) stands for \lq\lq thermal\rq\rq\
average
$$ M_j= \int^{ }_{ } D\varphi \ \varphi_ j {\rm e}^{-{\cal H}\{ \varphi\}} /Z
\eqno (1.6) $$

The Jacobian of the transformation is the determinant of the inverse
susceptibility matrix
$$ \left(\chi^{ -1} \right)_{ij}={\partial^ 2\Gamma \over \partial M_i\partial
M_j} \eqno (1.7) $$
and when a zero eigenvalue occurs it signals the inequivalence of the two
expansions and the occurrence of a transition. In Fourier transform $ \chi^{
-1}(q), $
and e.g., in zero momentum for a standard system, $ \chi^{ -1} $ vanishes in
zero field,
at $ T_c $ the Curie point
$$ \chi^{ -1} \left(q=0;T_c \right)=0\ , \eqno (1.8) $$
below which $ M\not= 0 $ even for $ H=0. $

This well known description of the paramagnetic to ferromagnetic transition
we would like now to extend to the random field system.
\vfill\eject
\vskip 24pt
\noindent {\bf 2. THE RANDOM FIELD ISING SYSTEM: Perturbation, a reminder}$
^{[ {\bf 20}]} $

Let us consider now $ H_i $ to be a quenched random field with a pure gaussian
probability distribution i.e.
$$ \eqalignno{ \charlvupup{ \dasharrowfill}{ H_i} & =0 & (2.1) \cr
\charlvupup{ \dasharrowfill}{ H_iH_j} & =\delta_{ ij}\Delta &  (2.2) \cr} $$
where the bar stands for probability average.
\vskip 17pt
{\bf 2.1 Direct averaging}

One may compute the $ H $ expansion of {\sl observables\/} and then perform
the Wick
average of the $ H $'s on each term of the expansion. In the {\sl
paramagnetic\/} phase,
to keep things simple, one obtains
$$ \eqalignno{  \charlvupup{ \dasharrowfill}{ W\{ H\}} &  = \sum^{ }_{ } {\rm
connected\ graphs\ with\ all\ pairs\ of\ } H {\rm ^\prime s} &  \cr  & {\rm \
} \ \ {\rm \ \ coalesced\ as\ in\ (2.2).} & (2.3) \cr} $$
$$ \eqalignno{ G(i;j) & \equiv \charlvupup{ \dasharrowfill}{ \left\langle
\varphi_ i\varphi_ j \right\rangle - \left\langle \varphi_ i \right\rangle
\left\langle \varphi_ j \right\rangle} = \sum^{ }_{ } connected\ {\rm graphs,\
rooted\ at\ } i\ {\rm and} \ j, &  \cr  &  \ \ \ \ \ \ \ \ \ \ \ \ \ \ \ \ \ \
\ \ \ \ \ \ \ \ \ \ \ \ \ {\rm with\ all\ pairs\ of\ } H {\rm ^\prime s\
coalesced.} & (2.4) \cr} $$
$$ \eqalignno{ C(i;j) & \equiv \charlvupup{ \dasharrowfill}{ \left\langle
\varphi_ i \right\rangle \left\langle \varphi_ j \right\rangle} = \sum^{ }_{ }
{\rm graphs\ made\ of\ two\ } disconnected\ {\rm pieces,\ respectively\
rooted} &  \cr  &  \ \ \ \ \ \ \ \ \ \ \ \ \ \ \ \ \ {\rm \ \ \ at\ } i\ {\rm
and} \ j\ {\rm with\ all\ pairs\ of\ } H {\rm ^\prime s\ colaesced.} & (2.5)
\cr} $$

The coalescence of all pairs of $ H $'s transforms the two disconnected pieces
into
one single {\sl field-connected\/} graph.
\vskip 17pt
{\bf 2.2 Averaging via the effective hamiltonian}

The above results can be recovered using the replica trick, that is,
computing
$$ \charlvupup{ \dasharrowfill}{ Z^n}= \charlvupup{ \dasharrowfill}{( {\rm
exp} \ W\{ H\})^ n}= {\rm exp} \left\{ n \charlvupup{ \dasharrowfill}{ W}+{n^2
\over 2} \left[ \charlvupup{ \dasharrowfill}{ W^2}- \left( \charlvupup{
\dasharrowfill}{ W} \right)^2 \right]+... \right\} \eqno (2.6) $$
where one then recovers the averaged free energy
$$ -F\equiv \charlvupup{ \dasharrowfill}{ W}= \left. \left( \charlvupup{
\dasharrowfill}{ Z^n}-1 \right) \right\vert_{ n \longrightarrow 0} \eqno (2.7)
$$
but also its successive fluctuation cumulants.

The effective hamiltonian is now,
$$ \eqalignno{{\cal H}_n & = \sum^{ }_ \alpha \left\{{ 1 \over 2} \sum^{ }_ p
\left(t_0+p^2 \right)\varphi^ \alpha (p)\varphi^ \alpha (-p)+{g \over 4!}
\sum^{ }_ j \left(\varphi^ \alpha_ j \right)^4 \right\} -{1 \over 2} \sum^{
}_{ \alpha ,\beta} \sum^{ }_ p\Delta \ \varphi^ \alpha (p)\varphi^ \beta (-p)
&  \cr  &   &  (2.8) \cr} $$

The propagator becomes a matrix $ {\bf G} $ with components
$$ G_{\alpha \beta} = \left\langle \varphi_ \alpha \varphi_ \beta
\right\rangle_ n- \left\langle \varphi_ \alpha \right\rangle_ n \left\langle
\varphi_ \beta \right\rangle_ n \eqno (2.9) $$
where the replica-thermal average is shown as $ \langle \rangle_ n. $ In the
paramagnetic phase
(no magnetization) $ \left\langle \varphi_ \alpha \right\rangle_ n=0. $ The
bare propagator $ G^0_{\alpha \beta} $ is the inverse of the
matrix $ \left(p^2+t_0 \right)\delta_{ \alpha \beta} -\Delta , $ i.e. with $
\left[G^0 \right]^{-1}=p^2+t_0 $
$$ G^0_{\alpha \beta} =G^0\delta_{ \alpha \beta} +{G^0\Delta G^0 \over
1-n\Delta G^0}\ . \eqno (2.10) $$
The first term is the {\sl connected\/} (bare) propagator, the last is the
{\sl field-connected\/} (bare) propagator (suppressing the $ H $-coalescence
into $ \Delta $'s it
falls into several disconnected pieces). In general we write
$$ {\bf G} \equiv G_{\alpha \beta} \equiv G_\alpha \delta_{ \alpha \beta}
+C_{\alpha \beta} \eqno (2.11) $$
Of course, in the paramagnetic region there is no explicit replica dependence
(in a RSB phase $ C_{\alpha \beta} $ however depends$ ^{[13-15]} $ upon the $
\alpha ,\beta $ overlap).

The observables calculated by direct averaging as in 2.1, are recovered via
$$ \eqalignno{ G & = \left.G_\alpha \right\vert_{ n \longrightarrow 0} &  \cr
C & = \left.C_{\alpha \beta} \right\vert_{ n \longrightarrow 0} & (2.12) \cr}
$$
Under RSB, one can relate$ ^{[14,15]} $ them by,
$$ C= \left.{1 \over n(n-1)} \sum^{ }_{ \alpha \not= \beta} C_{\alpha \beta}
\right\vert_{ n \longrightarrow 0}\ . \eqno (2.13) $$
\vskip 24pt
\noindent {\bf 3. LEGENDRE TRANSFORM}

To keep things simple we work from the paramagnetic phase i.e. with $
\left\langle \varphi_ \alpha \right\rangle_ n=0. $
We can then forget about the first Legendre transform that takes $ H $ into $
\left\langle \varphi_ \alpha \right\rangle_ n $
and concentrate on the second transform$ ^{[21-23]} $ taking $ \Delta_{ \alpha
\beta} (p) $ into $ \left\langle \varphi_ \alpha (p)\varphi_ \beta (-p)
\right\rangle_ n. $
We treat here $ \Delta $ as a source which we extend to values $ {\bf \Delta}
\equiv \Delta_{ \alpha \beta} (p) $ for
the purpose of generating appropriate observables, with in the end $ \Delta_{
\alpha \beta} (p) \longrightarrow \Delta . $
Together with $ W_n\equiv {\rm ln} \ \left( \charlvupup{ \dasharrowfill}{ Z^n}
\right) $ we introduce
$$ W_n\{ {\bf \Delta}\} =-\Gamma_ n\{ {\bf G}\} +{1 \over 2}\ {\rm tr} \ {\bf
\Delta} \ {\bf G} \eqno (3.1) $$
with
$$ G_{\alpha \beta} (p)={\partial W_n \over \partial \Delta_{ \alpha \beta}
(p)}\equiv G_\alpha (p)\delta_{ \alpha \beta} +C_{\alpha \beta} (p) \eqno
(3.2) $$
and
$$ \Delta_{ \alpha \beta} (p)={\partial \Gamma_ n \over \partial G_{\alpha
\beta} (p)} \eqno (3.3) $$
instead of (1.4,5).

The $ \Gamma_ n $ functional is itself given by
$$ -\Gamma_ n\{ {\bf G}\} ={1 \over 2}\ {\rm tr} \ {\rm ln} \ {\bf G} -{1
\over 2}\ {\rm tr} \ \left[G^0 \right]^{-1} {\bf G} +{\cal K}^{(1)}_{ }\{ {\bf
G}\} \eqno (3.4) $$
that is exhibiting components,
$$ \eqalignno{ -\Gamma_ n \left\{ G_\alpha ;C_{\alpha \beta} \right\} &  ={1
\over 2} \sum^{ }_ \alpha {\rm ln} \ G_\alpha +{1 \over 2}\ \doublelow{ {\rm
tr} \cr \alpha ,\beta \cr} \ {\rm ln} \ \left(\delta_{ \alpha \beta}
+G^{-1}_\alpha C_{\alpha \beta} \right) &  \cr  &  -{1 \over 2} \sum^{ }_
\alpha \left[G^0 \right]^{-1} \left[G_\alpha +C_{\alpha \alpha} \right]+
\sum^{ }_{ s=1}{\cal K}^{(1)}_s \left\{ G_\alpha ;C_{\alpha \beta} \right\} &
(3.5) \cr} $$
Here $ {\cal K}^{1)} $ is the 1-irreducible functional built with $ \varphi^
4_\alpha $ vertex and $ G_{\alpha \beta} $ lines
(i.e. such that by cutting off two such lines, whether connected $
\left(G_\alpha \delta_{ \alpha \beta} \right) $ or
field connected $ \left(C_{\alpha \beta} \right) $ the representative graph
does not fall into two
disconnected pieces). The subscript $ s $ in $ {\cal K}^{(1)}_s $ is the
number of free replica
indices, after account of the $ \delta_{ \alpha \beta} $ constraints of the
connected propagators.
\vskip 17pt
{\bf 3.1 Stationarity condition}

We consider separately, stationarity with respect to off diagonal and
diagonal components.
\vskip 12pt

{\sl Off diagonal component\/} $ \delta /\delta C_{\alpha \beta} : $
$$ \left[\uniset +G^{-1} {\bf C} \right]^{-1}_{\alpha \beta} G^{-1}_\beta
+\Delta_{ \alpha \beta} + \sum^{ }_{ s=2}{\delta{\cal K}^{(1)}_s \over \delta
C_{\alpha \beta}} =0 \eqno (3.6) $$
In the paramagnetic phase, $ C_{\alpha \beta} \longrightarrow C $ and, in the
above equation, only $ s=2 $
contributes
$$ G^{-1}CG^{-1}=\Delta +{\delta{\cal K}^{(1)}_2 \over \delta C}\{ G;C\} \eqno
(3.7) $$
\vskip 12pt

{\sl Diagonal component\/} $ \delta /\delta G_\alpha \equiv \delta /C_{\alpha
\alpha} : $

The equation obtained is more subtle to interpret because it contains {\sl
both\/}
connected and field-connected graphs and hence provides {\sl two\/} equations.
In
Appendix A it is shown that one equation is the Dyson equation for $ G_\alpha
$
$$ G^{-1}_\alpha - \left[G^0 \right]^{-1}+ \sum^{ }_{ s=1} \left[{\delta{\cal
K}^{(1)}_s \over \delta G_\alpha} \right]_{ {\rm conn}}=0 \eqno (3.8) $$
where, in the paramagnetic phase, only $ s=1 $ contributes. The other equation
is
the corresponding equation for $ C_{\alpha \alpha} $
$$ - \left[ \left[\uniset +G^{-1} {\bf C} \right]^{-1}G^{-1} {\bf C}
\right]_{\alpha \alpha} G^{-1}_\alpha +\Delta_{ \alpha \alpha} + \sum^{ }_{
s=2} \left[{\delta{\cal K}^{(1)}_s \over \delta G_\alpha} \right]_{f- {\rm
conn}}=0 \eqno (3.9) $$
{\sl Both\/} eqs.(3.6) and (3.9) can then be rewritten as (Appendix A)
$$ - \left[ \left[\uniset +G^{-1} {\bf C} \right]^{-1}G^{-1} {\bf C} G^{-1}
\right]_{\alpha \beta} +\Delta_{ \alpha \beta} + \sum^{ }_{ s=2}{\delta{\cal
K}^{(1)}_s \over \delta C_{\alpha \beta}} =0 \eqno (3.10) $$
an equation valid for $ \alpha \not= \beta $ and $ \alpha =\beta . $ This
seemingly formal result has the
consequence that, contrary to what happens in the standard SG
$$ C(x=1-\varepsilon ) \charlvmidlw{ \rightarrowfill}{ \varepsilon
\longrightarrow 0}C(1) \eqno (3.11) $$
showing that there is {\sl no jump\/} in a RSB phase as the $ \alpha \cap
\beta $ overlap $ x $ is taken to
be exactly equal to one\footnote{$ ^\ast $}{ Note that for consistency, the
extension $ \Delta \longrightarrow \Delta_{ \alpha \beta} , $
introduced here, has to satisfy a relationship analog to (3.11).}.
\vskip 17pt
{\bf 3.2 Second-derivative matrix}

We have
$$ {\cal M}_{\alpha \beta ;\gamma \delta} \left(p;p^{\prime} \right)={\delta
\Gamma_ n\{ {\bf G}\} \over \partial C_{\alpha \beta} (p)\partial C_{\gamma
\delta} \left(p^{\prime} \right)} \eqno (3.12) $$
a matrix in $ p,p^{\prime} $ and in replica pairs $ \alpha \beta ,\gamma
\delta . $ The structure in replica pair
space has been analyzed by de Almeida and Thouless$ ^{[24]} $ (for the
paramagnetic
region and in the absence of diagonal components of $ {\bf G} ). $ Here again
the
dangerous sector is the replicon one with the matrix
$$ \lambda_ R \left(p;p^{\prime} \right)=M_1-2M_2+M_3\equiv{\cal M}_R
\left(p;p^{\prime} \right) \eqno (3.13) $$
where
$$ \eqalignno{ M_1 & ={\cal M}_{\alpha \beta ;\alpha \beta} \left(p;p^{\prime}
\right) &  \cr M_2 & ={\cal M}_{\alpha \beta ;\alpha \gamma}
\left(p;p^{\prime} \right)={\cal M}_{\alpha \beta ;\gamma \beta}
\left(p;p^{\prime} \right) &  \cr M_3 & ={\cal M}_{\alpha \beta ;\gamma
\delta} \left(p;p^{\prime} \right) & (3.14) \cr} $$
The above expression simplifies greatly if one recognizes compensations
occurring between the $ M $ components. These compensations are
handily taken care of as follows.

Consider the first functional derivative
$$ \eqalignno{ \Delta_{ \alpha \beta} &  ={\delta \Gamma_ n \over \delta
C_{\alpha \beta}} &  (3.15) \cr \Delta_{ \alpha \alpha} &  ={\delta \Gamma_ n
\over \delta C_{\alpha \alpha}} ={\delta \Gamma_ n \over \delta G_\alpha} &
(3.16) \cr} $$
Contributing graphs are such that the ends $ \alpha ,\beta $ in $ \Delta_{
\alpha \beta} $ are necessarily
field-connected, whereas in $ \Delta_{ \alpha \alpha} $ the ends may be
connected (contributing to
the equation for $ G^{-1}_\alpha ) $ or field-connected (contributing to the
equation for $ C_{\alpha \alpha} ). $
Upon a second derivative, consider the connectedness of the {\sl new\/} end
points to
the pair of {\sl initial\/} end points. These may be connected or
field-connected.
The structure of eigenvalues (3.11-14) is such that, one recovers
$$ \eqalignno{ \lambda_ R \left(p;p^{\prime} \right) & = \left.M_1
\right\vert_{ {\rm conn}}\equiv \left.{\cal M}_{\alpha \beta ;\alpha \beta}
\left(p;p^{\prime} \right) \right\vert_{ {\rm conn}} & (3.17) \cr} $$
where the index conn. stands for the connectedness between the right and left
pairs.

All the graphs with field-connexions between the left and right pairs
compensate each other to only leave (3.17). From the explicit form of $
\Gamma_ n $
one gets, in exact form
$$ \lambda_ R \left(p;p^{\prime} \right)=G^{-1}_\alpha (p)G^{-1}_\beta
\left(p^{\prime} \right)\delta_{ p+p^{\prime} ;0} \left.- \sum^{ }_{
s=2}{\delta^ 2{\cal K}^{(1)}_s \over \delta C_{\alpha \beta} (p)\delta
C_{\alpha \beta} \left(p^{\prime} \right)} \right\vert_{ {\rm conn}} \eqno
(3.18) $$
where {\sl only\/} $ s=2 $ {\sl contributes in the \/}$ n \longrightarrow 0 $
{\sl limit.\/}

Note that $ \lambda_ R $ starts with an
{\sl attractive coupling\/} making it a candidate to come out with a null
eigenvalue.
\vskip 24pt
\noindent {\bf 4. }$ \Delta $-{\bf SUSCEPTIBILITY AND THE SG TRANSITION}

We are now in a situation that bears some analogy with the SG in field. In
the paramagnetic region we have $ C_{\alpha \beta} =C, $ the analog (now space
dependent) of
the SG order parameter $ q_{\alpha \beta} =q. $ As one crosses the line,
defined by the
vanishing of the lowest eigenvalue of $ \lambda_ R \left(p;p^{\prime} \right),
$ playing the role of the
Almeida-Thouless line, to avoid negative eigenvalues one has to break
replica-symmetry$ ^{[13-15]} $ and write $ C_{\alpha \beta} =C(x) $ where $
x=\alpha \cap \beta $ is the overlap of the replica pair (in
the Parisi$ ^{[16]} $ sense).

Just like in the pure system the vanishing of the jacobian signals the
occurrence of a singularity in the $ H $-susceptibility, here the vanishing of
the lowest $ \lambda_ R $ eigenvalue signals a singularity in the $ \Delta
$-susceptibility.
$$ {\cal G}_{\alpha \beta ;\gamma \delta} \left(p;p^{\prime} \right)\equiv{
\partial^ 2W_n \over \partial \Delta_{ \alpha \beta} (p)\partial \Delta_{
\gamma \delta} \left(p^{\prime} \right)} \eqno (4.1) $$
Indeed in its replicon sector, we have
$$ {\cal G}_R \left(p;p^{\prime} \right)={\cal G}_1 \left(p;p^{\prime}
\right)-2{\cal G}_2 \left(p;p^{\prime} \right)+{\cal G}_3 \left(p;p^{\prime}
\right) \eqno (4.2) $$
which is just the inverse of $ {\cal M}_R \left(p;p^{\prime} \right): $
$$ {\cal G}_R \left(p;p^{\prime} \right)= \left[{\cal M}_R \right]^{-1}
\left(p;p^{\prime} \right) \eqno (4.3) $$
that is
$$ {\cal G}_R \left(p;p^{\prime} \right)=G^2(p) \left[ \left.\delta_{
p+p^{\prime} ;0}+ \sum^{ }_{ p^{\prime\prime}}{\cal M}_{\alpha \beta ;\alpha
\beta} \left(p;p^{\prime\prime} \right) \right\vert_{ {\rm conn}}{\cal G}_R
\left(p^{\prime\prime} ;p^{\prime} \right) \right] \eqno (4.4) $$

One may also directly write out the standard SG susceptibility
$$ \chi_{ {\rm SG}} \left(r_1-r_2 \right)= \charlvupup{ \dasharrowfill}{
\left[ \left\langle \varphi \left(r_1 \right)\varphi \left(r_2 \right)
\right\rangle - \left\langle \varphi \left(r_1 \right) \right\rangle
\left\langle \varphi \left(r_2 \right) \right\rangle \right]^2} \eqno (4.5) $$
which is related to $ {\cal G}_R $ by
$$ \sum^{ }_{ p,p^{\prime}}{\cal G}_R \left(p;p^{\prime} \right)= \sum^{ }_{
1,2}\chi_{ {\rm SG}} \left(r_1-r_2 \right) \eqno (4.6) $$

It is striking to see that in the random field system, SG singularities are
confined to 2-replica contributions i.e. looking back at (2.6) into $
-F_2\equiv \charlvupup{ \dasharrowfill}{ W^2}- \left( \charlvupup{
\dasharrowfill}{ W} \right)^2 $
the free energy {\sl fluctuation\/} and not the free energy itself.

However, as soon as we are {\sl in a RSB phase, free energy fluctuations \/}$
F_2,F_3... ${\sl\
are no longer of order\/} $ n^2,n^3 $ respectively but all become proportional
to $ n $
and thus {\sl contribute in a finite way to the free energy\/}.

Let us see that effect on a simple example. Let us consider the lowest order
contribution to $ {\cal K}^{(1)}_2: $
$$ \sum^{ }_{ 1,2} \sum^{ }_{ \alpha ,\beta} C^4_{\alpha \beta} \left(r_1-r_2
\right) \eqno (4.7) $$
We have
$$ \eqalignno{ \sum^{ }_{ \alpha ,\beta} C^4_{\alpha \beta} &  = \sum^{ }_
\alpha \left[ \sum^{ }_{ \beta \not= \alpha} C^4_{\alpha \beta} +C^4_{\alpha
\alpha} \right] &  \cr  &  =n \left[- \int^{ 1-\varepsilon}_ 0 {\rm d} x\
C^4(x)+C^4(1) \right] &  \cr  &  =n \int^ 1_0x\ {\rm d} x\ { {\rm d} \over
{\rm d} x}\ C^4(x) & (4.8) \cr} $$
where one has used (3.11). Hence the $ O \left(n^2 \right) $ term is now $
O(n) $ and one
understands the vanishing of that contribution in the RS limit with the
vanishing of the derivative $ {\rm d} \ C(x)/ {\rm d} x. $ In general the
total number of derivatives (with
respect to overlaps $ x,y...) $ is equal to the number of replicas involves
minus
one $ (s-1 $ in $ {\cal K}^{(1)}_s). $ This is an unusual example where {\sl
the topology of the
graphs contributing to the free energy strongly depends on the phase one is
into.\/}
\vskip 24pt
\noindent {\bf 5. PHASE DIAGRAM}

To investigate, in the plane $ (\Delta ,T) $ what line is defined by the
occurrence
of a null eigenvalue, we first consider the Curie line. On that line the
progagators are massless and we write them as follows,
$$ \eqalignno{ C(p) & ={1 \over p^{2-\eta}} c \left({\Delta^ \omega \over
p^\theta} \right)\ . & (5.1) \cr} $$
Here we have $ \theta =2- \left(\bar \eta -\eta \right), $ $ \omega >0 $ $
(\omega =1 $ in the mean field limit), and $ c(y)=0 $ if $ y=0 $ (the pure
Ising
case), $ c(y)=y $ if $ y \longrightarrow \infty $ to recover a behavior in $
p^{-4+\bar \eta} . $ Hence we may take
$$ C(p)={\Delta^ \omega \over p^{4-\bar \eta}} \equiv{ \charlvupup{
\dasharrowfill}{ \Delta} \over p^{4-\bar \eta}} \ . \eqno (5.2) $$
As for the connected propagator we take first, for simplicity
$$ G(p)={1 \over p^{2-\eta}} \eqno (5.3) $$
noting that in the crossover region to the pure system, (5.3) will have to be
modified.
\vskip 12pt

{\sl (i)\nobreak\ Lowest order in \/}$ \bar \Delta : $ {\sl on the Curie
line\/}

To lowest order the eigenvalue equation reads,
$$ p^{4-2\eta} f_\lambda \left(\vec p \right)- \left(g\bar \Delta \right)^2
\int^{ }_{ }{ {\rm d}^Dp \over (2\pi )^D}C_2(q)f_\lambda  \left(\vec p-\vec q
\right)=\lambda f_\lambda \left(\vec p \right) \eqno (5.4) $$
with $ f_\lambda \left(\vec p \right) $ the eigenvector with $ \lambda $
eigenvalue and
$$ \eqalignno{ C_2(q) & = \int^{ }_{ }{ {\rm d}^Ds \over (2\pi )^D}{1 \over
s^{4-\bar \eta} \left(\vec s+\vec q \right)^{4-\bar \eta}} \equiv{ c \over
q^{8-D-2\bar \eta}} &  \cr c & ={1 \over (4\pi )^{D/2}}\Gamma \left(
\left(8-D-2\bar \eta \right)/2 \right){\Gamma^ 2 \left( \left(D-4+\bar \eta
\right)/2 \right) \over \Gamma \left(D-4+\bar \eta \right)} & (5.5) \cr} $$
Here $ C_2(q) $ can also be interpreted as the first cumulant contribution of
a
random temperature term.

To overcome the difficulty of solving the above integral equation (or in
Fourier transform, the \lq\lq Schr\"odinger\rq\rq\ equation with an \lq\lq
almost quartic\rq\rq\ kinetic
term) we resort to the Rayleigh-Ritz variational approach that provides an
{\sl upper bound,\/} by writing
$$ \eqalignno{ \lambda_ R & \equiv \lambda = \int^{ }_{ }{ {\rm d}^Dp \over
(2\pi )^D}\ p^{4-2\eta} \left\vert f \left(\vec p \right) \right\vert^ 2-
\left(g\bar \Delta \right)^2 \int^{ }_{ }{ {\rm d}^Dq \over (2\pi )^D}\
C_2(q)\phi \left(\vec q \right) & (5.6) \cr \phi (q) & = \int^{ }_{ }{ {\rm
d}^Dp \over (2\pi )^D}\ f^\ast \left(\vec p \right)f \left(\vec p-\vec q
\right) & (5.7) \cr} $$

Here $ f \left(\vec p \right) $ a {\sl normalized\/} trial wave function $
\langle {\rm i.e.} \ \phi (0)=1\rangle $ whose {\sl parameters\/} are
to be determined {\sl variationally.\/}

Since we are looking for the {\sl lowest\/} eigenvalue (\lq\lq zero-energy
bound state\rq\rq ) we
take $ f \left(\vec p \right)=f(p) $ and real. We can now {\sl scale\/} out
the unit length $ R. $
$$ \eqalignno{ \lambda &  ={a \over R^{4-2\eta}} -{ \left(g\bar \Delta
\right)^2c \over R^{2 \left(D-4+\bar \eta \right)}} \int^{ }_{ }{ {\rm d}^Dq
\over (2\pi )^D}\ {1 \over q^{8-D-2\bar \eta}} \phi (q) &  \cr  &  \equiv{ a
\over R^{4-2\eta}} - \left(g\bar \Delta \right)^2{b \over R^{2 \left(D-4+\bar
\eta \right)}} & (5.8) \cr} $$
Writing the stationarity condition with respect to $ R, $ one gets
$$ 0={(2-\eta )a \over \left(R^2 \right)^{2-\eta}} -{ \left(D-4+\bar \eta
\right) \left(g\bar \Delta \right)^2b \over \left(R^2 \right)^{D-4+\bar \eta}}
\eqno (5.9) $$
Hence solving from (5.8,9), we obtain finally
$$ \lambda = \left(D-6+\bar \eta +\eta \right) \left[{a \over \left(D-4+\bar
\eta \right) \left(R^2 \right)^{2-\eta}} \right]= \left(D-6+\bar \eta +\eta
\right) \left[{ \left(g\bar \Delta \right)^2 \over (2-\eta ) \left(R^2
\right)^{D-4+\bar \eta}} \right] \eqno (5.10) $$
with the length scale (i.e. correlation length)
$$ {1 \over R^2}= \left[{D-4+\bar \eta \over 2-\eta}{ b \over a} \left(g\bar
\Delta \right)^2 \right]^{{1 \over 6-D-\eta -\bar \eta}} \eqno (5.11) $$
As we rest on the Curie line we see that the eigenvalue {\sl upper bound\/}
remains
{\sl negative\/} for all $ \bar \Delta $ (except $ \bar \Delta^ \ast =0 $
where $ \lambda =0, $ a limit upon which we return
below).

It follows that, within the interval of dimension where the (ultraviolet) cut
off does not spoil the length scaling, the paramagnetic to ferromagnetic
transition is superceded by a paramagnetic to SG transition, provided it is
meaningful to keep only the lowest order contribution to the $ {\cal M}_R
\left(p;p^{\prime} \right) $ kernel.

The boundaries of the dimension interval are
$$ D_u=6-\eta \left(D_u \right)-\bar \eta \left(D_u \right)=6 \eqno  $$
and
$$ D_{\ell} =4-\bar \eta \left(D_{\ell} \right) \eqno  $$

Given that the standard dimensional reduction (and the associated $ \eta =\bar
\eta $ result)
is no more applicable, one is entitled to take $ \bar \eta =2\eta $ which is
correct$ ^{[25-27]} $
near $ D=2. $ The lower critical dimension $ D_{\ell} =2 $ then obtained by
using $ \theta =2- \left(\bar \eta -\eta \right) $
in the vicinity of $ D=2, $ that is $ \eta =1 $ for $ D=2. $

Thus, modulo the (inessential) changes that will be introduced below for a
treatment of the cross over region, what this very simple calculation is
telling us reduces to the following: the results obtained by perturbation to
all orders (dimensional reduction with $ \theta =2, $ and $ \bar \eta =\eta )
$ are superceded by the
occurrence of the SG transition which originates in the {\sl attraction\/}
existing
between pairs of distinct replicas. In contradistinction, and {\sl a
contrario\/}, for
\lq\lq animals\rq\rq\ (i.e. branched polymers) whose effective Lagrangean is
alike the
RFIM one but with a pure imaginary coupling$ ^{[28-30]}, $ the attraction
becomes a
{\sl repulsion,\/} and in a random field, dimensional reduction is indeed
correct$ ^{[29]}. $
\vskip 12pt

{\sl (ii)\nobreak\ \/}To test the robustness of the above result,
one may follow M\'ezard and Young$ ^{[14]} $ in adopting Bray's$ ^{[17]} $
approach
i.e. use a screened interaction for an $ m $-component system and work
consistenly to a given $ {1 \over m} $ order.

Eq.(5.6) is now replaced by
$$ \eqalignno{ \lambda &  = \int^{ }_{ }{ {\rm d}^Dp \over (2\pi )^D}\
p^{4-2\eta} f^2(p)-{ \left(g\bar \Delta \right)^2 \over m} \int^{ }_{ }{ {\rm
d}^Dq \over (2\pi )^D}\ S^2(q)C_2(q)\phi (q) &  \cr  &  -{ \left(g\bar \Delta
\right)^2 \over m} \int^{ }_{ }{ {\rm d}^Dq \over (2\pi )^D}\ S^2(q) \left[
\int^{ }_{ }{ {\rm d}^Dp \over (2\pi )^D}\ f(p){1 \over ( {\bf p} + {\bf q}
)^{4-\bar \eta}} \right]^2-O \left({1 \over m^2} \right) & (5.11) \cr} $$
with $ C_2(q), $ $ \phi (q) $ as of (5.5,7) and
$$ \eqalignno{ S(q) & =\mu^{ 6-D-\eta -\bar \eta} \left[1-\Theta \left(\mu
q_0-q \right) \right]+ \left({q \over q_0} \right)^{6-D-\eta -\bar \eta}
\Theta \left(\mu q_0-q \right) & (5.12) \cr q^{6-D-\eta -\bar \eta}_ 0 &
={g\bar \Delta \over (4\pi )^{D/2}}\Gamma \left( \left(6-D-\eta -\bar \eta
\right)/2 \right){\Gamma \left( \left(D-4+\bar \eta \right)/2 \right)\Gamma((
D-2+\eta) /2) \over \Gamma \left(D- \left({6-\bar \eta -\eta \over 2} \right)
\right)\Gamma \left({2-\eta \over 2} \right)\Gamma \left({4-\bar \eta \over 2}
\right)} &  \cr  &   &  (5.13) \cr} $$
Here $ \mu $ can either be unity, or chosen to take the best account of
screening. Again scaling out the unit length yields
$$ \eqalignno{ \lambda &  ={a \over R^{4-2\eta}} -{ \left(g\bar \Delta
\right)^2 \over m}{1 \over R^{2 \left(D-4+\bar \eta \right)}} \left\{ \int^{
\mu q_0R}_0{ {\rm d}^Dq \over (2\pi )^D}\ \left({q \over q_0R} \right)^{2
\left(6-D-\bar \eta -\eta \right)}A(q)+ \right. &  \cr  & \left. \int^{
\infty}_{ \mu q_0R}{ {\rm d}^Dq \over (2\pi )^D}\ \mu^{ 2 \left(6-D-\bar \eta
-\eta \right)D}A(q) \right\} -O \left({1 \over m^2} \right) & (5.14) \cr} $$
$$ A(q)=C_2(q)\phi (q)+ \left[ \int^{ }_{ }{ {\rm d}^Dp \over (2\pi )^D}\
f(p){1 \over ( {\bf p} + {\bf q} )^{4-\bar \eta}} \right]^2 \eqno (5.15) $$
Note that the $ R $-derivative with respect to integration boundaries does not
contribute. The stationarity condition upon $ R $ yields then
$$ \eqalignno{ O & =(2-\eta ) \left[{a \over R^{4-2\eta}} -{ \left(g\bar
\Delta \right)^2 \over m}{1 \over R^{2 \left(D-4+\bar \eta \right)}} \int^{
\mu q_0R}_0{ {\rm d}^Dq \over (2\pi )^D}\ \left({q \over q_0R} \right)^{2
\left(6-D-\bar \eta -\eta \right)}A(q) \right] &  \cr  &  -{ \left(g\bar
\Delta \right)^2 \over m}{ \left(D-4+\bar \eta \right) \over R^{2
\left(D-4+\bar \eta \right)}} \int^{ \infty}_{ \mu q_0R}{ {\rm d}^Dq \over
(2\pi )^D}\ \mu^{ 2 \left(6-D-\bar \eta -\eta \right)}A(q)\ . & (5.16) \cr} $$
leading again to the Rayleigh-Ritz approximation of the eigenvalue,
$$ \lambda ={ \left(D-6+\bar \eta +\eta \right) \over 2-\eta}{ \left(g\bar
\Delta \right)^2 \over m}{1 \over R^{2 \left(D-4+\bar \eta \right)}} \int^{
\infty}_{ \mu q_0R}{ {\rm d}^Dq \over (2\pi )^D}\ \mu^{ 2 \left(6-D-\bar \eta
-\eta \right)}A(q)\ . \eqno (5.17) $$
This expression remains negative for $ D<6, $ confirming the result obtained
in
(i).
\vskip 12pt

{\sl (iii)\nobreak\ \/}So far we have shown instability along the Curie line
when keeping
the terms in $ {1 \over m} \left(g_{ {\rm scr}}\bar \Delta \right)^2. $ One
would obtain analogous qualitative behavior for $ \left[{1 \over m} \left(g_{
{\rm scr}}\bar \Delta \right)^2 \right]^2 $
terms. The first {\sl repulsive\/} contribution only occurs as $ {g_{ {\rm
scr}} \over m}\bar \Delta \left[{ \left(g_{ {\rm scr}}\bar \Delta \right)^2
\over m} \right]^2. $
\vskip 12pt

{\sl (iv)\nobreak\ Lowest order in \/}$ \bar \Delta : ${\sl\ cross over
region\/}

If we want to use $ G(p) $ throughout the crossover to the pure limit $ \bar
\Delta =0, $ one
should replace (5.3) by
$$ G(p)={\bar \Delta^{ \left(\eta_ p-\eta \right)/\theta} \over p^{2-\eta}} g
\left({\bar \Delta^{ 1/\theta} \over p} \right) \eqno (5.18) $$
with $ g(x)\sim x^{\eta -\eta_ p} $ as $ x \longrightarrow 0 $ and $ g(x)\sim
C $ as $ x \longrightarrow \infty , $ the subscript $ p $ referring to the
pure limit, or alternatively, by
$$ G(p)={1 \over p^{2-\eta_ p}}\tilde g \left({\bar \Delta^{{ 1 \over \theta}}
\over p} \right) \eqno (5.19) $$
with $ \tilde g(0)=1 $ and $ \tilde g(x)\sim 1/x^{\eta -\eta_ p} $ as $ x
\longrightarrow \infty . $

Now Eq.(5.8) is replaced by
$$ \lambda ={\bar \Delta^{{ 2 \over \theta} \left(\eta -\eta_ p \right)} \over
R^{4-2\eta}} \int^{ }_{ }{ {\rm d}^Dp \over (2\pi )^D}p^{4-2\eta} f^2(p)\
g^{-2} \left(R\bar \Delta^{{ 1 \over \theta}} /p \right)-{\bar \Delta^ 2b
\over R^{2 \left(D-4+\bar \eta \right)}} \eqno (5.20) $$
and with the stationarity condition
$$ O={2-\eta \over D-4+\bar \eta} \cdot{\bar \Delta^{{ 2 \over \theta}
\left(\eta -\eta_ p \right)} \over R^{4-2\eta}} \cdot \left[ \int^{ }_{ }{
{\rm d}^Dp \over (2\pi )^D}p^{4-2\eta} f^2(p)\ g^{-2}(y/p) \left[1+{2 \over
2-\eta}{ y \over p}{\dot g(y/p) \over g(y/p)} \right] \right]-{\bar \Delta^ 2b
\over R^{2 \left(D-4+\bar \eta \right)}} \eqno (5.21) $$
where $ y\equiv R\bar \Delta^{{ 1 \over \theta}} , $ thus yielding, with
obvious notations,
$$ \lambda ={D-6+\bar \eta +\eta \over D-4+\bar \eta}{\bar \Delta^{{ 2 \over
\theta} \left(\eta -\eta_ p \right)} \over R^{4-2\eta}} \left[ \left\langle
g^{-2} \right\rangle +{2 \over 6-D-\bar \eta -\eta} \left\langle{ y \over
p}\dot gg^{-3} \right\rangle \right] \eqno (5.22) $$
A {\sl sufficient\/} condition to keep $ \lambda $ {\sl negative\/} is to have
$ g(x) $ be a {\sl monotonously
increasing\/} function, a very natural property given the above limiting
values
for $ x=0, x \longrightarrow \infty . $

The correlation length $ R $ is now given by (5.21) and one verifies that when
$ D $
is between $ D_{\ell} $ and $ D_u, $ $ \lambda $ vanishes with $ \bar \Delta .
$
\vskip 12pt

{\sl (v)\nobreak\ Lowest order in \/}$ \bar \Delta : $ {\sl away from the
Curie line\/}

As one is departing from the Curie line, the propagators become massive. In
Eq.(5.6) e.g., the \lq\lq kinetic\rq\rq\ contribution will be increased
slightly but the
\lq\lq potential\rq\rq\ one will be sharply decreased, the mass playing the
role of an
infrared cutoff. Hence the eigenvalue will increase and become null at some
point, on the SG transition line. To obtain that line we need the scaling
functions for the propagators which are now significantly more complex, since
they depend upon two variables $ x=\bar \Delta^{{ 1 \over \theta}} /p\equiv
\Delta_ 0/p $ and $ y=\delta T^{\nu_ p}/p $ where $ \delta T= \left\vert T-T_c
\left(\bar \Delta \right) \right\vert $ is
the distance, for a given $ \bar \Delta $ to the Curie line. If one is willing
to become
more speculative, one may use a typical scaling form,
restituting the appropriate behavior in all limits, as
$$ \eqalignno{ C^{-1}(p) & ={p^{4-\bar \eta} \over \Delta^{ 2-\bar \eta +\eta_
p}_0}+{(\delta T)^{\bar \gamma} \over \Delta^{ \left(\bar \gamma -\gamma_ p
\right)/\nu_ p}_0} &  \cr  &  =p^{2-\eta_ p} \left\{{ 1 \over x^{2-\bar \eta
+\eta_ p}}+x^{2-\eta_ p} \left({y \over x} \right)^{\bar \gamma /\nu_ p}
\right\} &  (5.23) \cr} $$
where $ \bar \gamma =\nu \left(4-\bar \eta \right), $ and
$$ \eqalignno{ G^{-1}(p) & =p^{2-\eta_ p}+p^{2-\eta} \Delta^{ \eta -\eta_
p}_0+{\delta T^\gamma \over \delta T^{\gamma -\gamma_ p}+\Delta^{ \left(\gamma
-\gamma_ p \right)/\nu_ p}_0} &  \cr  &  =p^{2-\eta_ p} \left\{ 1+x^{\eta
-\eta_ p}+y^{2-\eta_ p}{(y/x)^{ \left(\gamma -\gamma_ p \right)/\nu_ p} \over
1+(y/x)^{ \left(\gamma -\gamma_ p \right)/\nu_ p}} \right\} &  (5.24) \cr} $$
Leaving out a complicated discussion to be dealt with separately, let us just
consider the vicinity of the upper critical dimension.

Note first that the Curie line becomes now the locus of a SG/Ferro-SG
transition. However the $ G $ propagator being weakly dependent upon the SG
order
parameter, we shall assume it unchanged near $ D=6, $ that is given by
$$ T-T_c+a\Delta \simeq 0 \eqno (5.25) $$
where $ a $ is positive.

Let us compute the SG transition line for $ D=6-\varepsilon , $ at vanishing
values of $ \delta T $
and $ \Delta . $ Proceeding in the same manner as in (i) and (iv) we get to
leading
order
$$ \delta T\sim b\Delta^{ 4/\varepsilon} \eqno (5.26) $$
where $ b $ is positive and vanishes with $ \varepsilon . $ We thus have
$$ T-T_c\sim -a\Delta +b\Delta^{ 4/\varepsilon} \eqno (5.27) $$
i.e. the SG transition line starts {\sl tangent\/} to and remains {\sl very
close \/}to the
Curie line for small $ \varepsilon . $

As $ D $ decreases, the SG domain gets wider but too little is known about the
behaviour of the Curie line itself to decide whether there is reentrance
(i.e. whether the $ \Delta $ exponent of the $ b $ term in (5.27) can become
smaller than
the one of the $ a $ term).
\vskip 24pt

\noindent {\bf 6. CONCLUSION}

We have shown that for small enough values of $ g\bar \Delta , $ one obtains a
negative
upper bound for the eigenvalues of the inverse spin glass susceptibility, as
it occurs in the free-energy fluctuation.

This enforces the occurrence of a SG phase and the replica symmetry breaking
investigated by de Almeida \& Bruinsma, M\'ezard \& Young, M\'ezard \&
Monasson. It
enforces it for all $ D $ between the upper $ \left(D_u=6 \right) $ and lower
$ \left(D_{\ell} =2 \right) $ dimensions.
Thus the general properties obtained via perturbation to all orders$ ^{[2-4]},
$ for
$ D<6 $ (i.e. dimensional reduction with $ \theta =2) $ have no domain of
application on
the Curie line, if
the above results are not reversed for higher values of $ g_{ {\rm scr}}\bar
\Delta . $ Indeed, they
are then superceded by the SG transition for any $ D,\ 2<D<6. $

Whether this entails that the appropriate description is via a SG order
parameter as in M\'ezard-Young remains to be seen

(i)\nobreak\ although the above result is unlikely to be reversed, one would
like to
render foolproof the above derivation (by extending it to all orders in $
g\bar \Delta ). $

(ii)\nobreak\ this being given, it would also be desirable to see whether the
above SG
transition is not superceded itself by 3-replicas \lq\lq zero energy bound
states\rq\rq ,
4-replicas etc... (as contrasted with the 2-replica studied above).

Or put another way, just as we have seen above, that the SG transition
associated to the order parameter $ \left\langle \varphi_ \alpha \varphi_
\beta \right\rangle $ can be interpreted as governed by an
effective random temperature (i.e. mass, with a $ C_2(q)\sim 1/q^{
\left(8-D-2\bar \eta \right)} $
correlation), likewise one may ask whether higher order parameters e.g. $
\left\langle \varphi_ \alpha \varphi_ \beta \varphi_ \gamma \right\rangle $
etc... associated with higher random couplings$ ^{[31]} $ will not become
relevant
and spoil the above results.
\vskip 24pt
\noindent{\bf\ ACKNOWLEDGMENTS}

We thank T. Garel and M. M\'ezard for interesting discussions. One of us (CD)
also wishes to thank D. Fisher, T. Kirkpatrick, D. Thirumalai, A.P. Young for
fruitful exchanges, and D. Thirumalai and the IPST at the university of
Maryland for the
hospitality extended to him, under NSF contract CHE93-07884. Funding from the
BALATON program (\#94015) is also gratefully acknowledged.
\vfill\eject
\centerline{{\bf APPENDIX}}
\vskip 17pt

Here we show that the stationarity condition on the $ \Gamma_ n $ functional
(3.4,5)
yields eqs.(3.8-10) with the implication (3.11).

Consider the contributions associated with a given graph to $ {\cal K}^{(1)}\{
{\bf G}\} . $

 Then take a given choice for the $ {\bf G} $ components (with a constraint as
in $ G_\alpha \delta_{ \alpha \beta} $
or no constraint as in $ C_{\alpha \beta} ). $ The contributions of $ {\cal
K}^{(1)} $ are then associated
with $ {\cal K}^{(1)}_1,{\cal K}^{(1)}_2,...\ . $

Consider now the first functional derivative
$$ {\delta{\cal K}^{(1)}_s \over \delta G_\alpha} \eqno (A.1) $$
and the connectedness with respect to the input-output pair of lines (opened
by the functional derivation). Those two lines can be connected or
field-connected (if suppressing the $ \Delta $-coalescence the input-output
lines fall
apart) and one can always write
$$ {\delta{\cal K}^{(1)}_s \over \delta G_\alpha} = \left[{\delta{\cal
K}^{(1)}_s \over \delta G_\alpha} \right]_{ {\rm conn}}+ \left[{\delta{\cal
K}^{(1)}_s \over \delta G_\alpha} \right]_{ {\rm field-conn}} \eqno (A.2) $$
With (A.2) one can separate out the stationarity condition with respect to $
G_a $
(or $ C_{aa}) $ as (3.8) {\sl and\/} (3.9).

By inspection, one then writes
$$ \left[{\delta{\cal K}^{(1)}_s \over \delta G_\alpha} \right]_{ {\rm
field-conn}}= \left.{\delta{\cal K}^{(1)}_{s+1} \over \delta C_{\alpha \beta}}
\right\vert_{ \alpha =\beta} \eqno  $$
Noting that any $ \delta /\delta C_{\alpha \beta} $ contribution is by
definition field-connected. Hence
one obtains (3.10) and the result that the equation for $ C_{\alpha \alpha} $
(3.9) is obtained
by working with a pair of distinct replicas, and letting ({\sl after\/}
functional
derivation) $ \beta \longrightarrow \alpha , $ implying (3.11).
\vfill\eject
\centerline{{\bf REFERENCES}}
\vskip 17pt
\item{$\lbrack$1$\rbrack$} T. Natterman, J. Villain, Phase Transitions {\bf
11} (1988) 817.

\item{$\lbrack$2$\rbrack$} D.P. Belanger, A.P. Young, J. Magn. Magn. Mater.
{\bf 100} (1991) 272.

\item{$\lbrack$3$\rbrack$} A. Aharony, Y. Imry, S.K. Ma, Phys. Rev. Lett.
{\bf 37} (1976) 1364.

\item{$\lbrack$4$\rbrack$} A.P. Young, J. Phys. A {\bf 10} (1977) L257.

\item{$\lbrack$5$\rbrack$} G. Parisi, N. Sourlas, Phys. Rev. Lett. {\bf 43}
(1979) 744.

\item{$\lbrack$6$\rbrack$} Y. Imry, S.K. Ma, Phys. Rev. Lett. {\bf 35} (1975)
1399.

\item{$\lbrack$7$\rbrack$} J.Z. Imbrie, Phys. Rev. Lett. {\bf 53} (1984) 1747.

\item{$\lbrack$8$\rbrack$} A. Ochi, K. Watanabe, M. Kiyama, T. Shinjo, Y.
Bando, T. Takada,
J. Phys. Soc. Jpn. {\bf 50} (1981) 2777.

\item{$\lbrack$9$\rbrack$} H. Yoshizawa, D. Belanger, Phys. Rev. B {\bf 30}
(1984) 5220.

\item{$\lbrack$10$\rbrack$} C. Ro, G. Grest, C. Soukoulis, K. Levin, Phys.
Rev. B {\bf 31} (1985)
1681.

\item{$\lbrack$11$\rbrack$} T. Parker, A.E. Berkowitz, IEEE Trans. Magn. {\bf
25} (1989) 3647.

\item{$\lbrack$12$\rbrack$} U. Nowak, K.D. Usadel, Phys. Rev. B {\bf 44}
(1991) 7426.

\item{$\lbrack$13$\rbrack$} J.R.L. de Almeida, R. Bruinsma, Phys. Rev. B {\bf
35} (1987) 7267.

\item{$\lbrack$14$\rbrack$} M. M\'ezard, A.P. Young, Europhys. Lett. {\bf 18}
(1992) 653.

\item{$\lbrack$15$\rbrack$} M. M\'ezard, R. Monasson, Phys. Rev. B {\bf 50}
(1994) 7199.

\item{$\lbrack$16$\rbrack$} G. Parisi, J. Phys. A {\bf 13} (1980) 1887.

\item{$\lbrack$17$\rbrack$} A.J. Bray, Phys. Rev. Lett. {\bf 32} (1974) 1413.

\item{$\lbrack$18$\rbrack$} J. Yvon, Actualit\'es Sc. et Ind. {\bf 203}
(1935).

\item{$\lbrack$19$\rbrack$} See e.g. P. Morse and H. Feschbach \lq\lq Methods
of Theoretical
Physics\rq\rq , Mc Graw-Hill, New York (1952).

\item{$\lbrack$20$\rbrack$} See G. Parisi, Les Houches Session (1982), edited
by J.B. Zuber
and R. Stora (North Holland, Amsterdam (1984)).

\item{$\lbrack$21$\rbrack$} J. Luttinger, J. Ward, Phys. Rev. {\bf 118} (1960)
1417.

\item{$\lbrack$22$\rbrack$} C. De Dominicis, J. Math. Phys. {\bf 3} (1962)
983; {\bf 4} (1963) 255.

\item{$\lbrack$23$\rbrack$} C. De Dominicis, P.C. Martin, J. Math. Phys. {\bf
5} (1964) 17; 31.

\item{$\lbrack$24$\rbrack$} J.R.L. de Almeida, D.J. Thouless, J. Phys. A {\bf
11} (1978) 983.

\item{$\lbrack$25$\rbrack$} A.J. Bray, M.A. Moore, J. Phys. C {\bf 18} (1985)
L927.

\item{$\lbrack$26$\rbrack$} M. Schwartz, M. Gofman, T. Natterman, Physica A
{\bf 178} (1991) 6.

\item{$\lbrack$27$\rbrack$} M. Gofman, J. Adler, A. Aharony, A.B. Harris, M.
Schwartz, Phys.
Rev. Lett. {\bf 71} (1993) 1569.

\item{$\lbrack$28$\rbrack$} T. Lubenski, J. Isaacson, Phys. Rev. Lett. {\bf
41} (1978) 829.

\item{$\lbrack$29$\rbrack$} G. Parisi, N. Sourlas, Phys. Rev. Lett. {\bf 46}
(1981) 871.

\item{$\lbrack$30$\rbrack$} J. Fr\"ohlich, Les Houches Session XLIII (1984),
edited by K.
Osterwalder and R. Stora (North Holland, Amsterdam (1986)).

\item{$\lbrack$31$\rbrack$} D.S. Fischer, Phys. Rev. B {\bf 31} (1985) 7233.

\listrefs
\draftend
\end